\documentclass[12pt,a4paper]{cibb}

\usepackage{subfigure,graphicx}
\usepackage{amsmath,amsfonts,latexsym,amssymb,euscript,xr}
\usepackage{color,soul}

\title{\large $\ $\\ \bf Metabolic enrichment through functional gene rules}

\author{Davide Maspero$^{(1)}$, Claudio Isella $^{(2,3)}$, Marzia Di Filippo $^{(1,4)}$, Alex Graudenzi$^{(5)}$,  Sara Erika Bellomo$^{(2,3)}$, Marco Antoniotti$^{(5,6)}$, Giancarlo Mauri $^{(4,5)}$, Enzo Medico$^{(2,3)}$, Chiara Damiani$^{(4,5,*)}$}

\address{$\ $\\
(1) Department of Biotecnology and Bioscience, University of Milano-Bicocca, Milano, Italy
\\
\bigskip
(2) Candiolo Cancer Institute FPO IRCCS, Candiolo, Italy 
\\
\bigskip
(3) Department of Oncology, University of Torino, Torino, Italy
\\
\bigskip
(4) SYSBIO Centre of Systems Biology, Milan, Italy
\\
\bigskip
(5) Department of Informatics, Systems and Communication, University of Milano-Bicocca, Milan, Italy
\\
\bigskip
(6) Milan Center for Neuroscience
\\
\bigskip
(*) corresponding author (chiara.damiani@unimib.it)
}

\abstract{Metabolic network modeling, cross-sectional data, clustering, colorectal cancer, gene expression profile.
\\[17pt]
{\bf Abstract.} 
It is well known that tumors originating from the same tissue have different prognosis and sensitivity to treatments, depending on their molecular features. Over the last decade, cancer genomics consortia like the Cancer Genome Atlas (TCGA; https://cancergenome.nih.gov) have been generating thousands of cross-sectional data, spanning from genetic and epigenetic mutations to proteome profiles, for thousands of human primary tumors originated from various tissues. Thanks to the public access provided by such consortia to their datasets, it is today possible to analyze a broad range of relevant information such as gene sequences, expression profiles or metabolite footprints, to capture tumor molecular heterogeneity and improve patient stratification and clinical management.  To this aim, it is common practice to analyze datasets grouped into clusters based on clinical observations and/or molecular features. However, the identification of specific properties of each cluster that may be effectively targeted by therapeutic drugs still represents a challenging task. In this perspective, characterization of the metabolism of stratified patient cohorts may greatly help to select the best pharmacological treatment to prevent biomass formation and hence tumor growth. In this work, we define a method to generate an activity score for the metabolic reactions of different clusters of patients based on their transcriptional profile. This approach reduces the number of variables from many genes to few reactions, by aggregating transcriptional information associated to the same enzymatic reaction according to gene-enzyme and enzyme-reaction rules. As a proof of concept, we applied the methodology to a dataset of 244 RNAseq transcriptional profiles taken from patients with colorectal cancer (CRC), the second cause of cancer death in USA. CRC samples are typically divided into two sub-types: (i) tumors with microsatellite instability (MSI), associated with hyper-mutation and with CpG island methylation phenotype, and (ii) microsatellite stable (MSS) tumors, typically endowed with chromosomal instability. We report some key differences in the central carbon metabolism of the two clusters. We also show how the method can be used to describe the metabolism of individual patients and cluster them exclusively based on metabolic features.
}

\begin{document}
\thispagestyle{myheadings}
\pagestyle{myheadings}
\markright{\tt Preprint}

\section{\bf Scientific Background}



Genome-wide reconstructions of human metabolism, such as HMR \cite{mardinoglu2014genome} and Recon 2.2 \cite{swainston2016recon}, are today available.
%
%
These models include most metabolic reactions that may occur in a generic cell. To analyze a given tissue is necessary to extract a specific sub model with the subset of reactions that actually occur. Several methodologies  have been proposed to reach this goal using transcriptome, proteome or even metabolome data \cite{opdam2017systematic, machado2014systematic}. These methodologies require to set a threshold of gene expression or protein level whereby to decide whether keeping or removing a pathway. This partially arbitrary  parameter deeply affects the extracted model \cite{opdam2017systematic}. 

The above mentioned stoichiometric models are conceived to perform Flux Balance Analysis (FBA): a technique that exploits linear programming to compute the flux through each reaction under a steady state assumption \cite{orth2010flux}. The advantage of FBA is its use without knowledge about the enzymatic kinetic constants. FBA is based on the maximization (or minimization) of a given objective function. To simulate growth of tumor cells, maximization of biomass production is typically assumed. To perform a correct FBA it is necessary to define the nutritional constrains, i.e., the incoming nutrients and the outcoming products (exchange reactions). Unluckily this kind of data are yet rare in public databases. 

For these reasons, gene expression data are most often analyzed with gene set enrichment analysis algorithms. For instance, PANTHER classifies genes according to their function in several different ways: families and subfamilies are annotated with ontology terms \cite{mi2013large}. Nevertheless these algorithms do not account for the fact that metabolic reactions may be catalyzed by multi-domain enzymes or by different protein isoforms, as this would require knowledge on how genes are linked to enzymes. Association rules, which might now be retrieved from metabolic models, would provide more refined information than a mere list of genes. 

Along similar lines, metabolic reporter analyses, try to provide knowledge about variations in metabolite concentrations, starting from sets of genes classified according to the metabolite they associate with \cite{patil2005uncovering}. However these algorithms do not provide information about which reactions are up or down regulated, and thus on the putative targets for cancer treatment. Indeed reaction rates and metabolite concentrations are difficult to correlate.

We propose a methodology to infer deregulations of metabolic reactions, which, as opposed to FBA, does not need exometabolomic information nor optimality assumptions, but it simply takes as input transcriptomic profiles. For each reaction, we resolve the gene rules and compute a \textit{score} as a function of the transcript level. Thus it is not necessary to set a threshold \textit{a priori}.  We do not perform FBA simulations, but we only consider these scores as a static representation of the metabolic behavior of the given dataset. The methodology provides clear and useful information about which set of reactions are over (or under) expressed.


We apply the methodology to the manually reconstructed core model \textit{HMRcore} -- previously tested in \cite{damiani2017, DiFilippo2016} -- which contains pathways and reactions of central carbon metabolism extracted from the HMR genome-wide model \cite{mardinoglu2014genome} and the corresponding gene rules extracted from Recon 2.2 \cite{swainston2016recon}. 

\section{\bf Materials and Methods}

In this, work we started from the HMRcore model and from a dataset of colorectal cancer (CRC) downloaded from TCGA \cite{cancer2012comprehensive}. For the sake of completeness, we included in the model mitochondrial palmitate degradation and gluconeogenesis. In total we have 229 reactions with rules and 375 metabolic genes associate to them. We took gene rules from Recon2.2, where genes are identified with \textit{HGNC ID}  provided by the HUGO Gene Nomenclature Committee. 
Because each dataset identifies genes with different IDs we coded a script that automatically convert \textit{HGNC} to any other ID. 

The dataset under study contains level 3 RNAseq-v2 of CRC tissue samples gathered from 244 patients. Each tumor sample was already classified as MSS or MSI. We relied on this supervised classification to characterize with our method the two tumor subtypes.

We represented our  dataset in the from of a $n^g \times n^p$ matrix G, where $n^g$ is the total number of genes and $n^p$ is the number of patients. Each element $G_{g,p}$ is the RPKM (reads per kilobase per milion mapped reads) of gene $g$ in patient $p$ which is a proxy for the transcript level. We retain only the rows of this matrix corresponding to genes that are associated to reactions in the HMRcore model and used this new matrix as input for our algorithm.

Gene-enzyme rules are logical formulas that define the gene products a given reaction is catalyzed by and their relationship. These relationships are described by the logical operators \textit{AND} and \textit{OR}. The former defines set of genes that transcribe for different enzymatic sub units, which are all necessary for catalytic activity. The latter defines instead genes that transcribe for protein isoforms of the same enzyme, so that either one of them is sufficient to catalyze the reaction. 
For example the succinate-CoA ligase enzyme is formed by the subunit alpha (gene SUCLG1) and beta gene (SUCLG2) and catalyzes the reaction $Pi + succinyl$-$CoA + GDP 1 \leftrightarrow CoA + succinate + GTP$. The gene-enzyme rule for this reaction is therefore: SUCLG1 \textit{AND} SUCLG2. Conversely, ACACA and ACACB are respectively fully functional enzyme for the reaction \textit{acetyl-CoA carboxylase}, thus the rule is ACACA \textit{OR} ACACB. 

These logical operators can of course be combined to depict multi protein catalytic complexes or more complex situations involving both subunits and isoforms. For instance, ribonucleotide reductase is formed by two subunits: the catalytic (M1) and the regulatory one. The latter exists in two isoforms (M2 and M2B). The rule for this enzyme will therefore be RRM1 \textit{AND} (RRM2 \textit{OR} RRM2B).


To avoid the definition of a threshold and to be able to associate a continuous value to each reaction in different clusters, our method does not resolve the logical expressions in a Boolean fashion. We rather sum the transcripts of genes associated with the $OR$ operator, whereas we take the minimum transcript of genes associated with the operator \textit{AND} by respecting the precedence of the two operators. We thus obtain a score for each reaction in each patient.


%

The final output is therefore a $n^r \times n^p$ matrix $S$ where each element  $S_{r,p}$ is the score computed for reaction $r$ in patient $p$.

We split matrix S into two sub-matrixes according to tumor subtype classification. For each reaction we then test if the distribution of its scores is significantly different in the two clusters -- with the non parametric two-sample Kolmogorov-Smirnov test -- with a p-value threshold of 0.05. For every reaction that passed the test, we calculated the log2 fold change of the average score of that reactions in the two clusters. Because KS-test considers as significantly different distributions with the same mean but different standard deviation, we consider as relevant only log2 fold change greater than 0.263 (corresponding to a 20\% variation in the average score).

\section{\bf Results}
\begin{figure}[h]
	\vspace{3mm}
	\begin{center}
		\includegraphics[trim=0cm 0cm 0cm 3cm, width=0.9\linewidth]{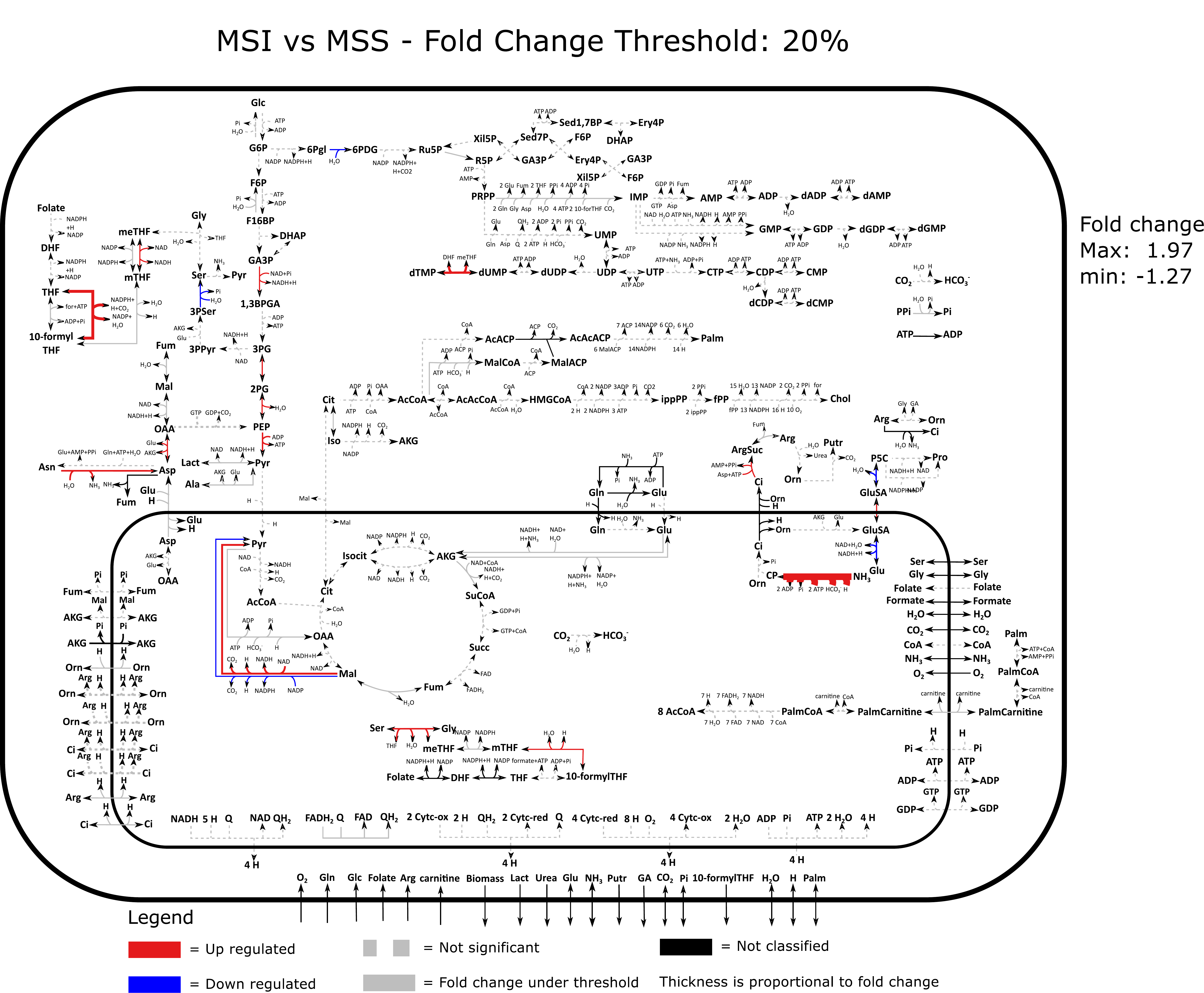}
		\caption {Map of HMRcore with fold change of reaction scores drawn on it. Red arrows refer to reactions upregulated in MSI; whereas blue arrows refer to reactions upregulated in MSS.  Black arrows refer to Not Classified reactions, i.e., reactions without information about the corresponding gene-enzyme rule. Dashed gray arrows refer to non significant deregulations according Kolgorov-Smirnof test. Solid gray arrows refer to reactions with a log2 fold change below 0.263}\label{fig:msivsmssfc20perc}
	\end{center}
	\vspace{-8mm}
\end{figure}

We applied our method to explain, in terms of metabolic behavior, the differences between MSS tumors, which typically show worse prognosis, and less common MSI tumors,  showing tumour-infiltrating lymphocytes and a better prognosis. We took the gene expression profiles and computed the reactions score as described in the previous section. To better visualize the results of our analyses, we drew the fold change of each reaction on the map in Fig.\ref{fig:msivsmssfc20perc}. 

The pathways more or less expressed in MSI with respect to MSS can be easily observed. For example three reactions in glycolysis pathway have a higher score in MSI respect to MSS. When comparing the two colorectal cancer subtypes (MSI versus MSS) we  indeed observed that the former is more glycolytic and, remarkably, it prefers to use NAD+ as cofactor to catalyze malate dehydrogenase reaction, whereas MSS prefers the NADP+ dependent reaction.

It can be observed that most reactions are not significantly different, suggesting that sharp metabolic differences in central carbon metabolism do not stand out, as long as this classification of patients is considered. On the one hand, our tool could be used to evaluate which existing partitions better highlight such differences. On the other hand, it could be directly used to identify new cluster of patients that maximize the metabolic differences among them, by performing an unsupervised cluster analysis on the scores that have been calculated for each patient. 

\begin{figure}[h]
	\vspace{3mm}
	\begin{center}
		\includegraphics[trim=0cm 0cm 0cm 3cm, width=0.9\linewidth]{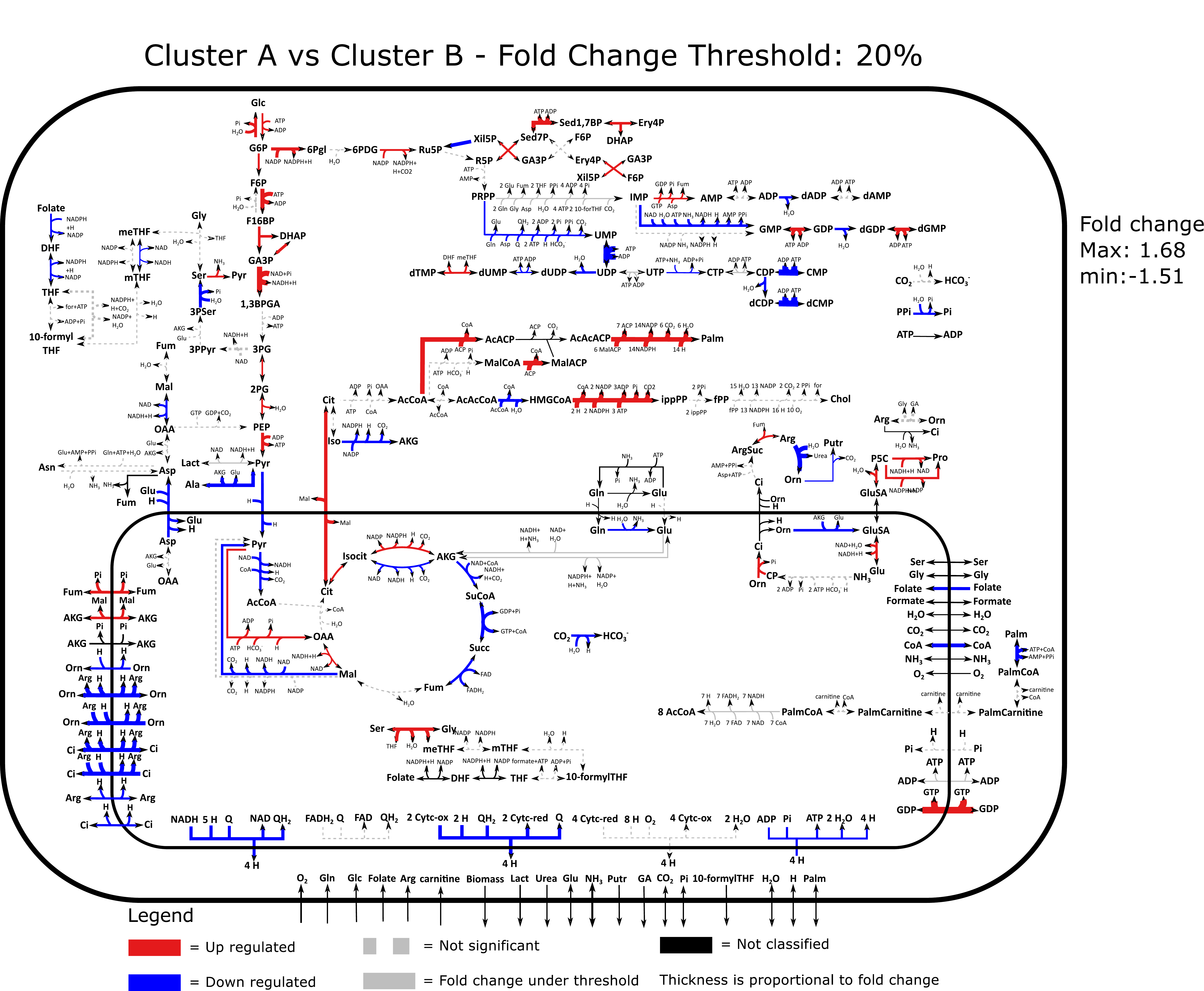}
\caption {Map of HMRcore with fold change of reactions scores drawn on it. Red arrows refer to reactions up-regulated in cluster A; whereas blue arrows refer to reactions up-regulated in cluster B. Black arrows refer to Not Classified reactions, i.e., reactions without information about the corresponding gene-enzyme rule. Dashed gray arrows refer to non significant deregulations according Kolgorov-Smirnof test. Solid gray arrows refer to reactions with a log2 fold change below 0.263}\label{fig:kmc20perc}

	\end{center}
	\vspace{-8mm}
\end{figure}

As a first approximation, we executed a k-means clustering on the reactions score obtained with our algorithm in order to split the dataset into two clusters ($k=2$). We refer to these new clusters as A and B. Based upon the new obtained classification, we repeated the procedure described before. As expected, the new obtained map (Fig.\ref{fig:kmc20perc}) shows greater differences between the two metabolisms. Interestingly, it is evident that one group with respect to the other shows a metabolism with more highly expressed glycolysis, fatty acid production from a reductive metabolism of glutamine, but a less expressed electron transport chain and oxidative phosphorylation. We remark that even though glycolysis is higher, the mitochondrial uptake of pyruvate is less expressed, indicating a lower oxidation of glucose, which is probably fermented to lactate.

It is worth noticing that although the two clusters have different cardinality (29 and 215 for cluster A and B respectively) the Kolmogorov-Smirnov hypothesis test should not be affected. Indeed, our results are confirmed even when we repeated the analysis considering the same cardinality for the two groups, i.e., by randomly selecting only 29 members from cluster B.  

In order to highlight the novelty of the results than can be obtained with our algorithm, we also performed the widely used Gene Set Enrichment Analysis (\textit{GSEA}) on the same dataset. We considered each reaction as a different gene set that includes the genes associated with that reaction. We considered as phenotypes \textit{MSS} and \textit{MSI}, so the input was exactly the same used before. We used a desktop application \cite{subramanian2007gsea} to run the analysis. \textit{GSEA} couldn't extract any significant result probably because a lot of get sets contain just one or few genes. All in all, our method is able to obtain more information and with a higher resolution as compared with a \textit{GSEA} on the same dataset.

\section{\bf Conclusion}
We introduced a method that computes a \textit{score} for each metabolic reaction, starting from transcriptome profiles. The quality of our results depends on the correctness of the gene-enzyme rules. As we took them from Recon2.2 as they were, our results might be improved by manually curating such rules.

In conclusion, our method is able to extract important information starting from any transcriptomics dataset, by reducing the dataset dimensionality from thousand genes to a smaller number of reactions.  The analysis focuses on the metabolic genes encompassed in a given metabolic network model.  We limited the analysis to central carbon pathways. In the next future we will extend the analysis to a genome wide model in order to show all the metabolic differences between tumors.

Finally, by clustering the scores obtained from clinical data, our method paves the way to easily linking particular prognosis or tumor aggressiveness to different metabolic traits. 


\section*{\bf Acknowledgments}
This work is supported with FOE funds from the Italian Ministry of Education, Universities and Research (MIUR, http://www.istruzione.it/) - within the Italian Roadmap for ESFRI Research Infrastructures - to GM and by grants from AIRC (investigator grant IG 16819 and 9970-2010 Special Program Molecular Clinical Oncology 5x1000) - to EM.

\bibliographystyle{apalike}
{\fontsize{10}{10}\selectfont

\end{document}